\documentclass[twoside]{ilcws08}
\usepackage[latin1]{inputenc}
\usepackage[dvips]{graphicx,epsfig,color}
\usepackage{wrapfig,rotating}
\usepackage{amssymb,amsmath,array}

\begin{document}
\title{
%%%%   Paper title goes here  %%%%%%%%%%%%%%
Latest Beam Test Results of the FONT4 ILC Intra-train Feedback System Prototype} %% 
%***********************************************************************
% AUTHORS INFORMATION AREA
%***********************************************************************
\author{P.~N.~Burrows$^1$, R. Apsimon$^1$, G.B. Christian$^2$, C. Clarke$^1$, 
B. Constance$^1$, H. Dabiri Khah$^1$, 
\\
T. Hartin$^1$, A.~Kalinin$^3$,
C. Perry$^1$,                    
J. Resta Lopez$^1$ and C. Swinson$^1$
\vspace{.3cm}\\
% Addresses and institutions (remove "1- " in case of a single institution)
%1- John Adams Institute, Oxford University, Oxford, OX1 3RH, UK
1- John Adams Institute at Oxford University, Oxford, OX1 3RH, UK
%% Remove the next three lines in case of a single institution
\vspace{.1cm}\\
%2- Second Author's Institution - Department \\
%Address of Second Author's Institution - Country\\
2- ATOMKI, Debrecen, Hungary
\vspace{.1cm}\\
3- Daresbury Laboratory, Warrington, UK
}
%%***********************************************************************
% END OF AUTHORS INFORMATION AREA
%***********************************************************************

\maketitle

\begin{abstract}
We present the design and preliminary results of a prototype beam-based digital feedback system for the Interaction Point of the %%@
International Linear Collider. A custom analogue front-end processor, FPGA-based digital signal processing board, and kicker drive %%@
amplifier have been designed, built, and tested on the extraction line of the KEK Accelerator Test Facility (ATF). The system was %%@
measured to have a latency of approximately 140 ns. 
\end{abstract}

\section{Introduction}

\begin{wrapfigure}{r}{0.5\columnwidth}
%\vspace{-0.4cm}
\centerline{\includegraphics[width=0.5\columnwidth,]{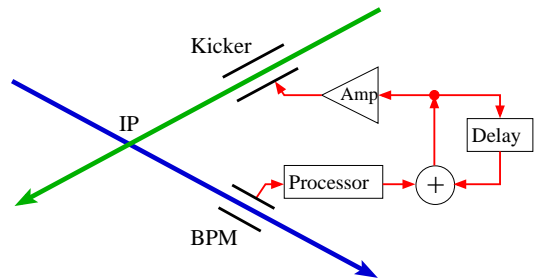}}
\vspace{-0.2cm}
\caption{Schematic of IP intra-train feedback system with a crossing angle. The deflection of the outgoing beam is registered in a %%@
BPM and a correcting kick applied to the incoming other beam.}\label{fbschematic}
\end{wrapfigure}

A number of fast beam-based feedback systems are required at the International electron-positron Linear Collider (ILC)~\cite{ILC}. %%@
At the interaction point (IP) a very fast system, operating on nanosecond timescales within each bunchtrain, is required to %%@
compensate for residual vibration-induced jitter on the final-focus magnets by steering the electron and positron beams into %%@
collision. A pulse-to-pulse feedback system is envisaged for optimising the luminosity on timescales corresponding to 5 Hz. Slower %%@
feedbacks, operating in the 0.1 - 1 Hz range, will control the beam orbit through the Linacs and Beam Delivery System. 

The key components of each such system are beam position monitors (BPMs) for registering the beam orbit; fast signal processors to %%@
translate the raw BPM pickoff signals into a position output; feedback circuits, including delay loops, for applying gain and %%@
taking account of system latency; amplifiers to provide the required output drive signals; and kickers for applying the position %%@
(or angle) correction to the beam. A schematic of the IP intra-train feedback is shown in Figure~\ref{fbschematic}, for the case %%@
in which the beams cross with a small angle; the current ILC design incorporates a crossing angle of 14 mrad.

Critical issues for the intra-train feedback performance include the latency of the system, as this affects the number of %%@
corrections that can be made within the duration of the bunchtrain, and the feedback algorithm. Previously we have reported on %%@
all-analogue feedback system prototypes in which our aim was to reduce the latency to a few tens of nanoseconds, thereby %%@
demonstrating applicability for `room temperature' Linear Collider designs with very short bunchtrains of order 100ns in length, %%@
such as NLC, GLC and CLIC~\cite{CLIC}. We achieved total latencies (signal propagation delay + electronics latency) of 67ns %%@
(FONT1)~\cite{FONT1}, 54ns (FONT2)~\cite{FONT2} and 23ns (FONT3)~\cite{FONT3}. 

We report the latest results on the design, development and beam testing of an ILC prototype system that incorporates a digital %%@
feedback processor based on a state-of-the-art Field Programmable Gate Array (FPGA). The use of a digital processor will allow for %%@
the implementation of more sophisticated algorithms which can be optimised for possible beam jitter scenarios at ILC. However, a %%@
penalty is paid in terms of a longer signal processing latency due to the time taken for digitisation and digital logic %%@
operations. This approach is now possible for ILC given the long, multi-bunch train, which includes parameter sets with c. %%@
3000/6000 bunches separated by c. 300/150ns respectively.  Initial results were reported previously~\cite{FONT4}.

\section{FONT4 Design}

\begin{wrapfigure}{r}{0.5\columnwidth}
%\centerline
{\includegraphics
[width=0.9\columnwidth]
{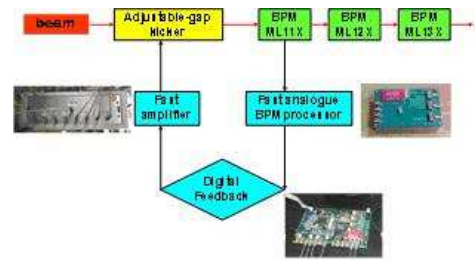}}
\vspace{-7cm}
\caption{Schematic of FONT4 at the ATF extraction beamline showing the relative locations of the kicker, BPMs and the elements of %%@
the feedback system.}\label{atfschematic}
\end{wrapfigure}

A schematic of the FONT4 feedback system prototype and the experimental configuration in the ATF extraction beamline is shown in %%@
Figure~\ref{atfschematic}. The layout is functionally equivalent to the ILC intra-train feedback system. An upstream dipole %%@
corrector magnet can be used to steer the beam so as to introduce a controllable vertical position offset in stripline BPM ML11X. %%@
The BPM signal is initially processed in a front-end analogue signal processor. The analogue output is then sampled, digitised and %%@
processed in the digital feedback board to provide an analogue output correction signal. This signal is input to a fast amplifier %%@
that drives an adjustable-gap stripline kicker~\cite{feather}, which is used to steer the beam back into nominal vertical %%@
position. BPMs ML12X and ML13X can serve as independent witnesses of the beam position. 

The ATF damping ring can be operated so as to provide an extracted train that comprises 3 bunches separated by an interval that is %%@
tuneable in the range 140 - 154 ns. This provides a short ILC-like train which can be used for controlled feedback, or %%@
feed-forward~\cite{FF} system tests. 

FONT4 has been designed as a bunch-by-bunch feedback with a latency goal of less than 140ns. This meets the minimum ILC %%@
specification of c. 150ns bunch spacing, as well as the smallest spacing allowed at ATF. This will allow measurement of the first %%@
bunch position and correction of both the second and third ATF bunches. The correction to the third bunch is important as it %%@
allows test of the 'delay loop' component of the feedback, which is critical for maintaining the appropriate correction over a %%@
long ILC bunchtrain.

\begin{wrapfigure}{r}{0.5\columnwidth}
\centerline
{\includegraphics[width=0.45\columnwidth]{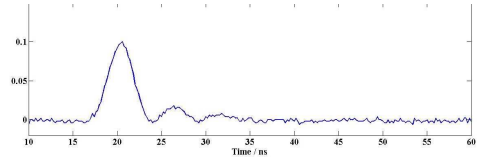}}
\caption{Voltage output of BPM signal processor.}\label{bpmvoltage}
\end{wrapfigure}

The design of the front-end BPM signal processor is based on that for FONT3. The top and bottom (y) stripline BPM signals were %%@
added and subtracted using a hybrid, to form a sum and difference signal respectively. The resulting signals were band-pass %%@
filtered and down-mixed with a 714 MHz local oscillator signal which was phase-locked to the beam. The resulting baseband signals %%@
are low-pass filtered. The hybrid, filters and mixer were selected to have latencies of the order of a few nanoseconds, in an %%@
attempt to yield a total processor latency in the range 5-10ns. The output pulse is illustrated in Figure~\ref{bpmvoltage}. The %%@
measured latency is around 10ns~\cite{FONT4}. 

\begin{wrapfigure}{r}{0.5\columnwidth}
%\centerline
{\includegraphics[width=0.9\columnwidth]{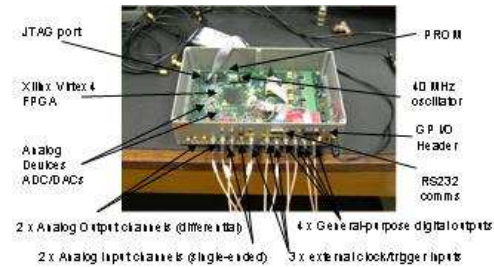}}
\vspace{-7cm}
\caption{FONT4 digital feedback board.}\label{font4board}
\end{wrapfigure}

The custom digital feedback processor board is shown in Figure~\ref{font4board}. There are two analogue signal input (output) %%@
channels in which digitisation is performed using Analog Devices ADCs (DACs) which can be clocked at up to 105 (210) Ms/s. The %%@
digital signal processing is based on a Xilinx Virtex4 FPGA which can be clocked at up to 500MHz. The FPGA is clocked with a 357 %%@
MHz source, derived from the ATF master oscillator and hence locked to the beam. Logic operations are triggered with a pre-beam %%@
signal. The ADC/DAC are clocked at 357/4 Ms/s. The analogue BPM processor output signal is sampled at the peak to provide the %%@
input  signal to the feedback.

The gain stage is implemented via a lookup table stored in FPGA RAM, alongside the reciprocal of the sum signal for charge %%@
normalisation. The delay loop is implemented as an accumulator on the FPGA. The output is converted back to analogue and used as %%@
input to the driver amplifier.

The driver amplifier was manufactured by TMD Technologies~\cite{tmd}, a UK-based RF company. The amplifier was specified to %%@
provide +-30A of drive current into the kicker~\cite{feather}, whose striplines were shorted at the upstream end (nearer the %%@
incoming beam). The risetime, starting at the time of the input signal, was specified as 35ns to reach 90\% of peak output. The %%@
output pulse length was specified to be up to 10 microseconds. Although current operation is with only 3 bunches in a train of %%@
length c. 300ns, it is planned in future to operate ATF with extracted trains of 20 or 60 bunches with similar bunch spacing; the %%@
design allows for this upgrade.

\section{Beam Test Results} 

\begin{wrapfigure}{r}{0.5\columnwidth}
%\vspace{-5cm}
%\centerline
{\includegraphics[width=0.7\columnwidth]{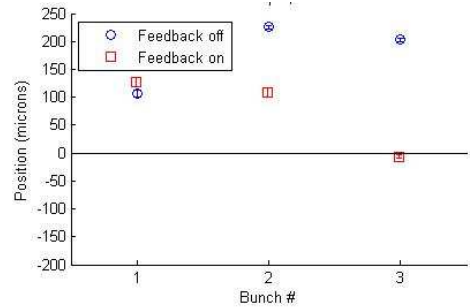}}
%\caption{Beam position vs. bunch number: low (top), medium (middle) and high (bottom) gain.}\label{fbresultsa}
%\end{wrapfigure}
%
%\begin{wrapfigure}{r}{0.5\columnwidth}
%\centerline
%\vspace{-11.5cm}
%{\includegraphics[width=0.7\columnwidth]{fbresults2c.eps}}
%\caption{Beam position vs. bunch number: low (top), medium (middle) and high (bottom) gain.}\label{fbresultsa}
%\end{wrapfigure}
%
%\begin{wrapfigure}{r}{0.5\columnwidth}
%\centerline
%\vspace{2cm}
%{\includegraphics[width=0.7\columnwidth]{fbresults2b.eps}}
\vspace{-4cm}
\caption{Beam position vs. bunch number.}\label{fbresults}
\end{wrapfigure}

We report the preliminary results of the most recent beam tests, performed in May 2008.  The beam was steered successively into %%@
different vertical positions spanning a range of about +-200 microns centred around nominal zero. An example is shown in %%@
Figure~\ref{fbresults}. The data shown are the averages over 10 or 11 pulses. The feedback is seen to under- and over-correct for %%@
low and high gain, respectively, and it is possible to make a 'perfect' correction of bunch 3 by an appropriate gain choice. It %%@
should be noted, however, that the bunchtrain is not straight, there being a sagitta of order 100 microns between the incoming %%@
bunches, and this complicates the operation of the feedback system. Nevertheless the system works as expected.

\begin{wrapfigure}{r}{0.5\columnwidth}
%\centerline
{\includegraphics[width=0.9\columnwidth]{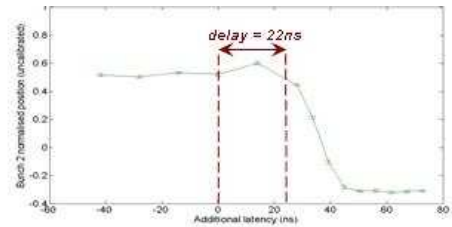}}
\vspace{-7.5cm}
\caption{Latency measurement.}\label{latency}
\end{wrapfigure}

The latency was measured by deliberately delaying the kick to bunch 2, and observing the kick vs. added delay %%@
(Figure~\ref{latency}). This used a special version of the FPGA firmware, without the inclusion of real-time charge normalisation. %%@
The delay at which bunch 2 stops being kicked, defined as 90\% of full kick to bunch 2, corresponds to a latency equal to the %%@
bunch spacing.  The difference between the 90\% kick point and zero added delay gives a measure of the amount of timing slack in %%@
the system, and hence, subtracting this from the bunch spacing of 154 ns yields a latency of c. 132 ns. The inclusion of real-time %%@
charge-normalisation uses an extra 3 clock cycles in the FPGA firmware, increasing the system latency by 8.4ns to c.140ns.

\section{Acknowledgments}
This work is supported by the Commission of the European Communities under the 6$^{th}$ Framework Programme "Structuring the %%@
European Research Area", contract number RIDS-011899, and by the UK Science and Technology Facilities Council.

%\verb?\section*{Appendix A}?

%\section{Bibliography}

% ****************************************************************************
% BIBLIOGRAPHY AREA
% ****************************************************************************

\begin{footnotesize}
% IF YOU DO NOT USE BIBTEX, USE THE FOLLOWING SAMPLE SCHEME FOR THE REFERENCES
% ----------------------------------------------------------------------------

\end{footnotesize}

% ****************************************************************************
% END OF BIBLIOGRAPHY AREA
% ****************************************************************************

\end{document}